\documentclass[12pt]{article}
\usepackage{multicol}
\usepackage{amssymb}
\usepackage{amsfonts,amssymb,amsmath,amsthm}
\usepackage{epsfig}
\usepackage{slashed}
\usepackage{graphicx}
\usepackage{float}
\usepackage{makecell}
\usepackage{subfig}
\usepackage{mathrsfs}
\usepackage{bbm}
\usepackage{cite}
\usepackage{enumerate}
\usepackage{longtable}
\usepackage{booktabs}
\usepackage[colorlinks,linkcolor=red,anchorcolor=red,citecolor=green]{hyperref}
\makeatletter 

\makeatletter

\newcommand{\Rmnum}[1]{\expandafter\@slowromancap\romannumeral #1@}
\makeatother

\begin{document}
\renewcommand{\thefootnote}{\fnsymbol{footnote}}
\begin{titlepage}

\vspace{10mm}
\begin{center}
{\Large\bf Quasinormal mode and stability of optical black holes in moving dielectrics}
\vspace{10mm}

{{\large Yang Guo${}^{}$\footnote{\em E-mail: guoy@mail.nankai.edu.cn}
and Yan-Gang Miao}${}^{}$\footnote{\em Corresponding author.}\footnote{\em E-mail: miaoyg@nankai.edu.cn}

\vspace{3mm}
${}^{}${\normalsize \em School of Physics, Nankai University, Tianjin 300071, China}
}
\end{center}

\vspace{5mm}
\centerline{{\bf{Abstract}}}
\vspace{6mm}
We study the quasinormal mode and stability of optical black holes in moving dielectrics. The results show that the real part of complex frequencies is inversely proportional to but the absolute value of an imaginary part is proportional to a refractive index.
We obtain the conditions for forming a black hole horizon in moving dielectrics. 
Moreover, we investigate the evolution behavior of optical black holes with respect to a refractive index at a high overtone number and find that an optical black hole undergoes one phase transition from an unstable mode to a stable one when the refractive index is big enough.

\vspace{5mm}
\noindent

\vspace{5mm}

\end{titlepage}

\newpage
\renewcommand{\thefootnote}{\arabic{footnote}}
\setcounter{footnote}{0}
\setcounter{page}{2}
\pagenumbering{arabic}
\vspace{1cm}

\section{Introduction}
Einstein founded the general relativity in 1915 and then predicted the existence of gravitational waves in 1916.
After one full century, the gravitational wave signal of a binary black hole merger was directly detected~\cite{GW} for the first time. The ringdown stage of black hole mergers carries the ``fingerprint" of black holes, i.e. the quasinormal mode frequency, which is determined only by the parameters of black holes, such as mass, charge and spin. As a result, the study of quasinormal modes gives a possibility to measure the parameters accurately, and provides~\cite{TGW,TNH,GVS} the test of general relativity and no-hair theorem as well.

From the prediction to the first direct detection, the gravitational wave was undoubtedly the greatest recognition to the exploration of gravity in the past 100 years. 
This progress indicates that it is very difficult to observe gravitational waves because the gravity is the weakest when compared to the other three interactions in experiments on Earth.
In addition, the information of gravity we can get now originated far away in the distant past. 
Consequently, theorists have been looking~\cite{UL,ULPP,ULP} for a medium that can be equivalent to the gravitational field in a laboratory. Such a treatment is also called `analogue gravity'~\cite{BLV}, with which one can get information in lab that is analogous to gravity. 
Many meaningful systems of `analogue gravity' have been constructed~\cite{MV,FLP}, such as sound waves in background fluids, moving media and slow light media in condensed matter physics. Correspondingly, an analogue of black holes was proposed~\cite{WGU,ULP}, where a moving dielectric could be equivalent to a gravitational field and the propagating of light through a dielectric vortex could lead to an Aharonov-Bohm phase shift~\cite{AB}. 
After investigating the optical path of propagation in a moving medium and changing the Gordon optical metric~\cite{WG,YOR} into a static one, De Lorenci, et al. suggested~\cite{DKO} an optical black hole in moving dielectrics. 
However, the metric given in ref.\cite{DKO} is not specified, which leaves the issue that some important properties of optical black holes, such the quasinormal mode frequency and the relation between frequencies and refractive indices, etc, cannot be uncovered.

Here we focus on the quasinormal modes of optical black holes.  For the details of perturbation theory of black holes, see, for instance, the reviews~\cite{KS,BCO,KZ}. When the background of spacetime of a black hole undergoes a perturbation, the black hole oscillates damply with a complex frequency, called a quasinormal mode, where its real part represents the oscillation of black holes and its imaginary part determines the scale of damping time. It is worth noting that the quasinormal  mode of black holes with high damping oscillations, i.e. with a large absolute value of imaginary parts, has a connection~\cite{JY,SH} to black hole thermodynamics, which is a kind of quantum  behaviors of black holes. However, the highly damped modes are often hard to be solved numerically. Fortunately, we can deal with such an issue for optical black holes and obtain the stable modes at high overtone numbers by adjusting the refractive index of media. That is, we discover that one phase transition from an unstable mode to a stable one occurs at high overtone numbers when the refractive index of optical black holes is big enough. Therefore, we have the opportunity to gain some insight into quantum effect of black holes through the analogue --- optical black holes.

The organization of this paper is as follows. In section~\ref{sec2}, we derive the Schr\"{o}dinger-like  equation of optical black holes and determine the range of refractive indices within which an event horizon can form. In section~\ref{sec3}, we compute the quasinormal mode frequencies numerically and analyze the phase transition at high overtone numbers. Finally, we conclude in section~\ref{sec4}.

\section{Formlism}\label{sec2}
The static metric describing the propagation of light rays in a moving medium with the four-velocity $u^{\mu}=\gamma(1,\beta,0,0)$ takes~\cite{DKO} a general form,
\begin{eqnarray}
\text{d}s^2=g_{00}\text{d}t^2-\left(g_{00}\right)^{-1}\text{d}r^2-
r^2d{\Omega}^2/\eta^2,\label{static} 
\end{eqnarray}
where $g_{00}=\gamma^2(1-\eta^2\beta^2)/(\mu\epsilon$), $\gamma=(1-\beta^2)^{-1/2}$, and $\eta=\pm\sqrt{\mu\epsilon}$ is the refractive index of media with the magnetic permeability $\mu$ and the dielectric permittivity $\epsilon$. 
A black hole horizon can be formed if the constraint $\eta^2\beta^2(r_h)=1$ is imposed~\cite{DKO}. For a spherically symmetric $\beta=\beta(r)$, we specify\footnote{Because $\beta$ is the ratio of the speed of light in a medium to the speed of light in the vacuum, which is less than one, so the radial coordinate just takes the values less than the maximum of horizon radii for a physical $\beta$. The maximal horizon radius equals one which corresponds to an infinite large refractive index of media. Therefore, our choice of $\beta$ function is reasonable since it leads to a natural consequence that $\beta$ is definable only inside an optical black hole. On the other hand, the radial coordinate can take values from the horizon radius to infinity in the metric because the metric depends on the squares of $\beta$ and $\gamma$. This characteristic of our optical black hole model is clearly shown in eq.~(\ref{metric}).} $\beta(r)=(1-r)^{1/2}$, which of course satisfies this constraint at the horizon.
In fact, it is not the only function we can choose but the most straightforward one. The reason is that this $\beta(r)$ function gives rise to an analogous Schwarzschild black hole, that is,  
the static metric given by eq.~(\ref{static}) can now be written in a Schwarzschild-like form,

\begin{eqnarray}
\text{d}s^2=\left(1-\frac{1-\eta^{-2}}{r}\right)\text{d}t^2-\left(1-\frac{1-\eta^{-2}}{r}\right)^{-1}\text{d}r^2-
r^2d{\Omega}^2/\eta^2.\label{metric}
\end{eqnarray}
Moreover, our choice of $\beta(r)$ function will lead to the range of refractive indices that coincides with optics of nonuniformly moving media, see the clarification in detail at the end of this subsection.  

As was known, the field strength tensor $F_{\mu\nu}$ is governed by the Maxwell equations,
\begin{eqnarray}
{F^{\mu\nu}}_{;\nu}=0,\label{Max}
\end{eqnarray}
where $F_{\mu\nu}\equiv A_{\nu,\mu}-A_{\mu,\nu}$ and $A_{\mu}$ is the electromagnetic vector potential. 
For the spherically symmetric case, one can expand~\cite{RTV} the vector potential by using the spherical harmonics function in a 4-dimensional vector form,
\begin{eqnarray}
A_{\mu}(t,r,\theta,\phi)=\sum_{l,m}\left(\begin{bmatrix}
0 \\
0 \\
\frac{a^{lm}(t,r)}{\sin\theta}\partial_{\phi}Y_{lm}(\theta,\phi)\\ 
-a^{lm}(t,r)\sin\theta\partial_{\theta}Y_{lm}(\theta,\phi)  \\ 
\end{bmatrix}
+\begin{bmatrix}
f^{lm}(t,r)Y_{lm}(\theta,\phi)\\
h^{lm}(t,r)Y_{lm}(\theta,\phi)\\
k^{lm}(t,r)\partial_{\theta}Y_{lm}(\theta,\phi)\\ 
k^{lm}(t,r)\partial_{\phi}Y_{lm}(\theta,\phi)  \\ 
\end{bmatrix}\right),\label{sep}
\end{eqnarray}
where the first term corresponds to the odd parity contribution of the field strength tensor $F_{\mu\nu}$,  
while the second term to the even parity contribution. 
 This parity classification is consistent with the Regge and Wheeler's gravitational perturbation~\cite{RW}. For the odd parity, $\Phi(t,r)=a^{lm}(t,r)$, while for the even parity $\Phi(t,r)=r^{2}l(l+1)(-i\omega h^{lm}-df^{lm}/dr)$, 
see refs.~\cite{RTV,VCL} for the details. 
Substituting eq.~(\ref{sep}) into the Maxwell equations eq.~(\ref{Max}), defining the tortoise coordinate  $\frac{dr_{*}}{dr}=\eta\left(1-\frac{1-\eta^{-2}}{r}\right)^{-1}$ in terms of eq.~(\ref{metric}), and making the separation of time and radial coordinate $\Phi(t,r)=e^{-i\omega t}\Phi(r)$,  we obtain the Schr\"odinger-like equation,

\begin{eqnarray}
\frac{d^2\Phi(r)}{dr_{*}^2}+[\omega^2-V(r)]\Phi(r)=0,
\end{eqnarray}
where the effective potential is given by 
\begin{eqnarray}
V(r)=\left(1-\frac{1-\eta^{-2}}{r}\right)\frac{l(l+1)}{r^2}.\label{EV}
\end{eqnarray}
We mention that $\omega$ is just the quasinormal mode frequency we shall calculate numerically.

\begin{figure}[H]
	\begin{center}
		\includegraphics[width=90mm]{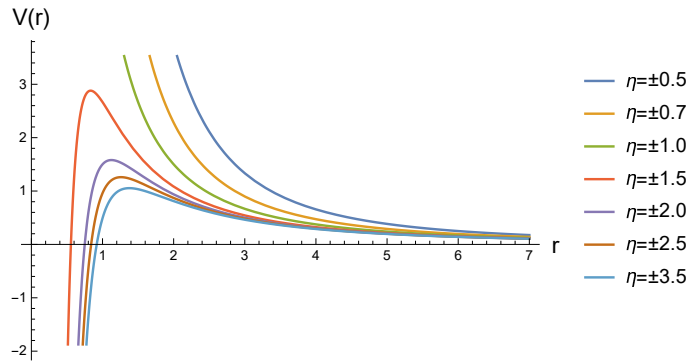}
	\end{center}
	\caption{$V(r)$ versus $r$ at the angular quantum number $l=2$ for seven different refractive indices, $\pm$0.5, $\pm$0.7, $\pm$1.0, $\pm$1.5, $\pm$2.0, $\pm$2.5, and $\pm$3.5, respectively, which are represented in different colors.}
	\label{V}
\end{figure}

We plot the dependence of potential $V$ on the radial coordinate $r$ for different values of refractive indices in Figure \ref{V}. Note that the potential depends only on the absolute value of refractive indices, which shows that a negative refractive index is possible in the optical black hole model described by eq.~(\ref{metric}). 

As can be seen from Figure~\ref{V}, the curves have no intersections on the horizontal axis when the absolute value of refractive indices is not larger than one, $\rvert \eta\rvert\leq1$, which means that  the boundary condition $V(r_{h})\rightarrow0$ is not satisfied in this case. As a result, we deduce that a black hole horizon can be formed only if the absolute value of refractive indices is larger than one, $\rvert \eta\rvert>1$. 
Here we make a remark that our choice of $\beta(r)$ function, $\beta(r)=(1-r)^{1/2}$, is appropriate because its corresponding range of refractive indices coincides with that suggested in ref.~\cite{ULP}, $1\leq \eta<\infty$, for a kind of optical black holes with metric solutions. The position of event horizons of metric solutions, $r_{h}=1-\eta^{-2}$,  will approach one asymptotically when the absolute value of refractive indices is going to infinity.
In particular, we emphasize that the negative index of refraction is possible and plays the same role as that of the positive index in our model of optical black holes.

\section{Result}\label{sec3}
According to the effective potential eq.~(\ref{EV}), we can compute the corresponding quasinormal mode frequencies numerically. We utilize the efficient and stable methods~\cite{YH,MO,KZZ,MT} that have brought a lot of attention recently.
One method we adopt is based on the Borel-Pad$\acute{\text e}$ summation~\cite{YH}.
The other method we use is the so-called higher order WKB-Pad$\acute{\text e}$ approach suggested first in ref.~\cite{MO} and developed later in refs.~\cite{KZZ,MT}. The two methods greatly improve the stability and accuracy of approximate calculations based on the WKB approach~\cite{WKB}.



\subsection{Quasinormal mode frequency}\label {sec:lsmall}
In this subsection we  are going to use the Borel-Pad$\acute{\text e}$ summation and the higher order WKB-Pad$\acute{\text e}$ approach, respectively,  to compute the quasinormal mode frequencies. We take the three different values of refractive indices, $\eta=5$, $\eta=10$, and $\eta=15$, respectively.
The real parts are presented in Table \ref{tab1} and the imaginary parts in Table \ref{tab2}. 
In Table \ref{tab1}, the real parts of frequencies in both the geometric unit and Hz unit are provided, and in Table \ref{tab2}, besides the imaginary parts in the geometric unit, the absolute values of their reciprocals are provided in millisecond, where the black hole mass is taken, for instance, to be ten times of the solar mass.
The main purpose to add the Hz unit 
and the millisecond unit 
for frequencies is only to provide an intuitive understanding of optical black holes when one compares to the black holes in Universe. 

\begin{center}

	\begin{longtable}{ccccc}
		
		\caption{The real parts of quasinormal mode frequencies of optiacl black holes }	
		
		\label{tab1} \\	
		\hline
		\hline
		{\small$\eta=5$ }& \qquad{ \small  Borel-Pad$\acute{\text e}$}& { \qquad\small $f$ (Hz)}&\qquad {\small  WKB-Pad$\acute{\text e}$}&{ \qquad\small $f$ (Hz)}\\ 	
		\hline
		$l=1$ $n=0$     &\qquad  0.515152 &\qquad 830.144&\qquad    0.517215 & \qquad 833.469\\
		$l=1$ $n=1$   	&\qquad  0.435540 &\qquad 701.853&\qquad    0.446822 &\qquad 720.034\\
		$l=2$ $n=0$		&\qquad  0.952235 &\qquad 1534.48&\qquad    0.953324 &\qquad 1536.24\\
		$l=2$ $n=1$		&\qquad  0.901456 &\qquad 1452.66&\qquad    0.909463 &\qquad 1465.56\\
		$l=2$ $n=2$     &\qquad  0.818274 &\qquad 1318.61&\qquad    0.835802 &\qquad1346.86\\
		$l=3$ $n=0$  	&\qquad  1.367787 &\qquad 2204.13&\qquad    1.368539 &\qquad 2205.34\\
		$l=3$ $n=1$		&\qquad  1.330811 &\qquad 2144.54&\qquad    1.336953 &\qquad 2154.44\\
		$l=3$ $n=2$		&\qquad  1.264246 &\qquad 2034.28&\qquad    1.278817 &\qquad 2060.76\\
		$l=3$ $n=3$		&\qquad  1.179851 &\qquad 1901.28&\qquad    1.203996 &\qquad1940.19\\
		\hline
		\hline
		{\bf\small$\eta=10$ }&\qquad { \small Borel-Pad$\acute{\text e}$}&\qquad {\small $f$ (Hz)}&\qquad {\small  WKB-Pad$\acute{\text e}$}&\qquad {\small $f$ (Hz)}\\ 
		\hline
		$l=1$ $n=0$     &\qquad   0.501109 &\qquad 807.515&\qquad   0.501542 &\qquad808.212\\
		$l=1$ $n=1$   	&\qquad   0.430745 &\qquad 694.126&\qquad   0.433282 &\qquad698.214\\
		$l=2$ $n=0$		&\qquad   0.924207 &\qquad 1489.32&\qquad   0.922435 &\qquad 1486.46\\
		$l=2$ $n=1$		&\qquad   0.880225 &\qquad 1418.44&\qquad   0.881904 &\qquad1421.15\\
		$l=2$ $n=2$     &\qquad   0.806446 &\qquad 1299.55&\qquad   0.810475 &\qquad1306.04\\
		$l=3$ $n=0$  	&\qquad   1.326910 &\qquad 2138.26&\qquad   1.327069 &\qquad2138.51\\
		$l=3$ $n=1$		&\qquad   1.295217 &\qquad 2087.18&\qquad   1.296439 &\qquad2089.15\\
		$l=3$ $n=2$		&\qquad   1.236915 &\qquad 1993.23&\qquad   1.240065 &\qquad1998.31\\
		$l=3$ $n=3$		&\qquad   1.161963 &\qquad 1872.45&\qquad   1.167511 &\qquad1881.39\\ 
		\hline
		\hline
		{\small$\eta=15$ }&\qquad { \small Borel-Pad$\acute{\text e}$}&\qquad{\small $f$ (Hz)}&\qquad{\small  WKB-Pad$\acute{\text e}$}&\qquad{\small $f$ (Hz)}	\\ 
		\hline
		$l=1$ $n=0$     &\qquad  0.498557 &\qquad803.402&\qquad 0.498743 &\qquad803.702  \\
		$l=1$ $n=1$   	&\qquad  0.429800 &\qquad692.603&\qquad 0.430864 &\qquad694.318  \\
		$l=2$ $n=0$		&\qquad  0.919179 &\qquad1481.22&\qquad 0.919277 &\qquad1481.37  \\
		$l=2$ $n=1$		&\qquad  0.876262 &\qquad1412.06&\qquad 0.876982 &\qquad1413.22  \\
		$l=2$ $n=2$     &\qquad  0.804191 &\qquad1295.92&\qquad 0.805952 &\qquad1298.76  \\
		$l=3$ $n=0$  	&\qquad  1.319595 &\qquad2126.47&\qquad 1.319663 &\qquad2126.58  \\
		$l=3$ $n=1$		&\qquad  1.288686 &\qquad2076.66&\qquad 1.289205 &\qquad2077.50  \\
		$l=3$ $n=2$		&\qquad  1.231784 &\qquad1984.96&\qquad 1.233145 &\qquad1987.16  \\
		$l=3$ $n=3$		&\qquad  1.158570 &\qquad1866.98&\qquad 1.160996 &\qquad1870.89  \\ 
		\hline
		\hline

	\end{longtable}
\end{center}

\begin{center}

	\begin{longtable}{ccccc}
	
	\caption{The imaginary parts of quasinormal mode frequencies of optical black holes}	
	
	\label{tab2} \\	
		  \hline
		  \hline
		     {\small$\eta=5$ }& \qquad{ \small  Borel-Pad$\acute{\text e}$}& \qquad{\small $\tau$ (ms)}&\qquad {\small  WKB-Pad\'e}&\qquad{\small $\tau$ (ms)}\\ 	
		     \hline
$l=1$ $n=0$     &\qquad - 0.176049i &\qquad3.525&\qquad - 0.192683i &\qquad 3.221\\
$l=1$ $n=1$   	&\qquad - 0.568868i &\qquad1.091&\qquad - 0.611720i &\qquad 1.014\\
$l=2$ $n=0$		&\qquad - 0.181265i &\qquad3.423&\qquad - 0.197926i &\qquad 3.135\\
$l=2$ $n=1$		&\qquad - 0.558867i &\qquad1.110&\qquad - 0.605646i &\qquad 1.025\\
$l=2$ $n=2$     &\qquad - 0.973612i &\qquad0.637&\qquad - 1.044974i &\qquad 0.594\\
$l=3$ $n=0$  	&\qquad - 0.182520i &\qquad3.400&\qquad - 0.199200i &\qquad 3.115\\
$l=3$ $n=1$		&\qquad - 0.555387i &\qquad1.112&\qquad - 0.603601i &\qquad 1.028\\
$l=3$ $n=2$		&\qquad - 0.949099i &\qquad0.654&\qquad - 1.025138i &\qquad 0.605\\
$l=3$ $n=3$		&\qquad - 1.371611i &\qquad0.452&\qquad - 1.471522i &\qquad 0.422\\
            \hline
            \hline
            {\bf\small$\eta=10$ }&\qquad { \small Borel-Pad$\acute{\text e}$}&\qquad {\small  $\tau$ (ms)}&\qquad {\small   WKB-Pad\'e}&\qquad {\small $\tau$ (ms)}\\ 
            \hline
$l=1$ $n=0$     &\qquad - 0.182927i &\qquad3.392&\qquad - 0.186844i &\qquad3.321\\
$l=1$ $n=1$   	&\qquad - 0.582776i &\qquad1.065&\qquad - 0.593183i &\qquad1.046\\
$l=2$ $n=0$		&\qquad - 0.188009i &\qquad3.301&\qquad - 0.191928i &\qquad3.233\\
$l=2$ $n=1$		&\qquad - 0.575995i &\qquad1.077&\qquad - 0.587293i &\qquad1.057\\
$l=2$ $n=2$     &\qquad - 0.995858i &\qquad0.623&\qquad - 1.013308i &\qquad0.612\\
$l=3$ $n=0$  	&\qquad - 0.189242i &\qquad3.280&\qquad - 0.193164i &\qquad3.213\\
$l=3$ $n=1$		&\qquad - 0.573780i &\qquad1.082&\qquad - 0.585310i &\qquad1.060\\
$l=3$ $n=2$		&\qquad - 0.975598i &\qquad0.636&\qquad - 0.994073i &\qquad0.624\\
$l=3$ $n=3$		&\qquad - 1.402497i &\qquad0.442&\qquad - 1.426931i &\qquad0.435\\ 
            \hline
            \hline
            {\small$\eta=15$ }&\qquad { \small Borel-Pad$\acute{\text e}$}&\qquad{\small  $\tau$ (ms)}&\qquad{\small  WKB-Pad\'e}&\qquad{\small  $\tau$ (ms)}	\\ 
            \hline
$l=1$ $n=0$     &\qquad - 0.184079i &\qquad3.371&\qquad - 0.185801i &\qquad3.340  \\
$l=1$ $n=1$   	&\qquad - 0.585314i &\qquad1.060&\qquad - 0.589873i &\qquad1.052  \\
$l=2$ $n=0$		&\qquad - 0.189135i &\qquad3.281&\qquad - 0.190857i &\qquad3.251  \\
$l=2$ $n=1$		&\qquad - 0.579026i &\qquad1.071&\qquad - 0.584016i &\qquad1.063 \\
$l=2$ $n=2$     &\qquad - 0.999929i &\qquad0.621&\qquad - 1.007653i &\qquad0.616 \\
$l=3$ $n=0$  	&\qquad - 0.190362i &\qquad3.260&\qquad - 0.192086i &\qquad3.231 \\
$l=3$ $n=1$		&\qquad - 0.576961i &\qquad1.076&\qquad - 0.582044i &\qquad1.066  \\
$l=3$ $n=2$		&\qquad - 0.980358i &\qquad0.633&\qquad - 0.988526i &\qquad0.628  \\
$l=3$ $n=3$		&\qquad - 1.408152i &\qquad0.441&\qquad - 1.418968i &\qquad0.437  \\ 
\hline
\hline

\end{longtable}
\end{center}

The above results show that the imaginary parts of quasinormal mode frequencies keep negative when $l\ge n$, where $l$ is angular quantum number and $n$ is overtone number, which indicates the stability of systems. 
The second and fourth columns in Table~\ref{tab1} and Table~\ref{tab2} are the frequencies calculated in terms of the 50th order Borel-Pad$\acute{\text e}$ summation and the 13th order WKB-Pad\'e approach, respectively, and the first seven digits are maintained.
In ref.~\cite{YH}, the results obtained by the Borel-Pad$\acute{\text e}$ summation and Leaver's method~\cite{EWL} were compared and the consistency of the two methods was determined. Here we can see, e.g from Table~\ref{tab1}, that the values obtained by the Borel-Pad$\acute{\text e}$ summation agrees well with that by the WKB-Pad$\acute{\text e}$ method 
when the refractive index is not less than 5 and $l$ is greater than $n$. 
For example, we take two sets of data, ($\eta=5$, $l=3$, $n=3$) and ($\eta=15$, $l=3$, $n=0$), respectively, and find that the two real parts, 1.179851 and 1.203996 in the first set, differ by $2.01\%$, and 1.319595 and 1.319663 in the second set, just by $0.07\%$. As a result, the results of the Borel-Pad$\acute{\text e}$ summation and the WKB-Pad\'e method show an excellent agreement in the case of high refractive indices and $l$ greater than $n$.

Note that the frequencies are given in the geometric units~\cite{KS}. Thus we convert, for an intuitive understanding of optical black holes, the unit of real parts into Hz and the unit of absolute values of the reciprocals of imaginary parts into millisecond, and 
list the real parts in the third and fifth columns of Table~\ref{tab1} and the absolute values of the reciprocals of imaginary parts in the third and fifth columns of Table~\ref{tab2}, respectively. 
For instance, the quasinormal mode frequency is 0.515152-0.176049i by the Borel-Pad$\acute{\text e}$ summation in the case of ($\eta=5$, $l=1$, $n=0$), 
which is equivalent to an oscillation frequency 830.144Hz and a damping time 3.525 ms for a black hole with 10 times of the solar mass. 

\begin{figure}[H]
	\centering
	\includegraphics[width=60mm]{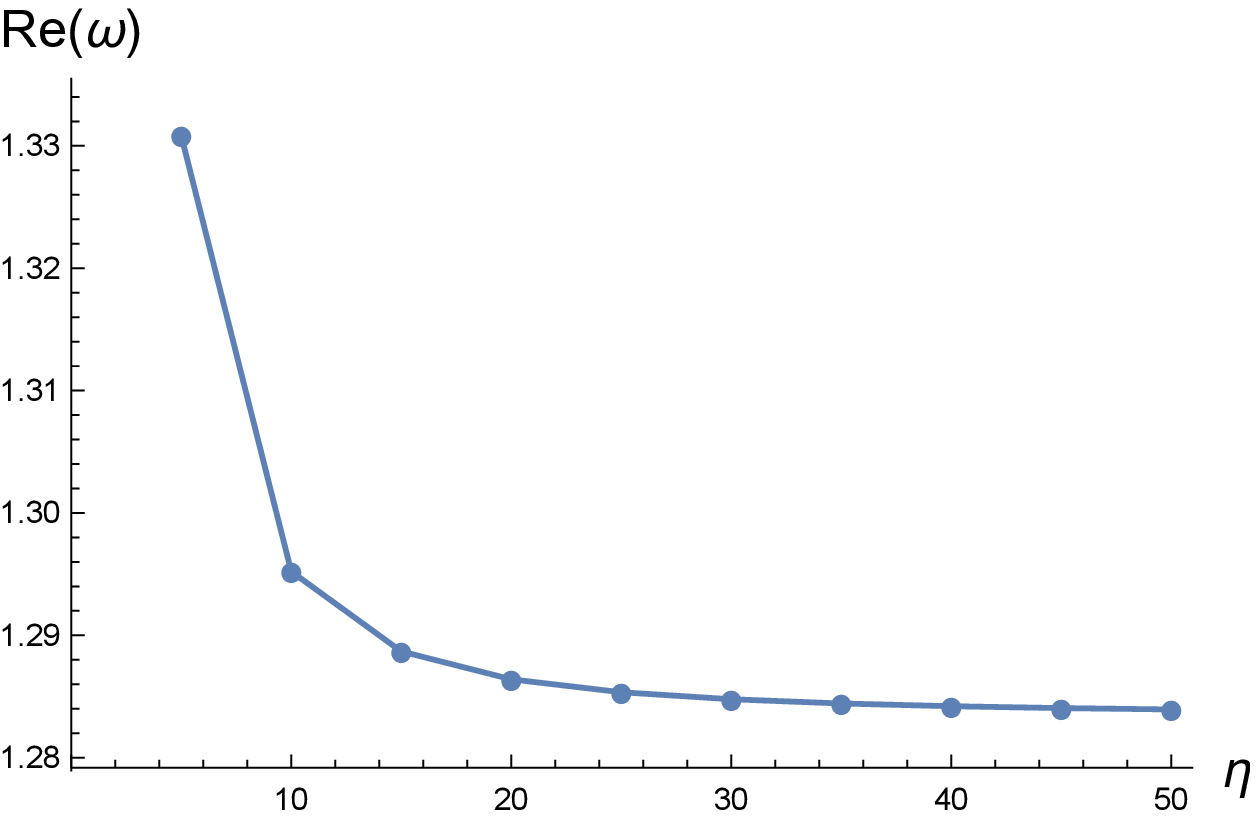}  \qquad \includegraphics[width=60mm]{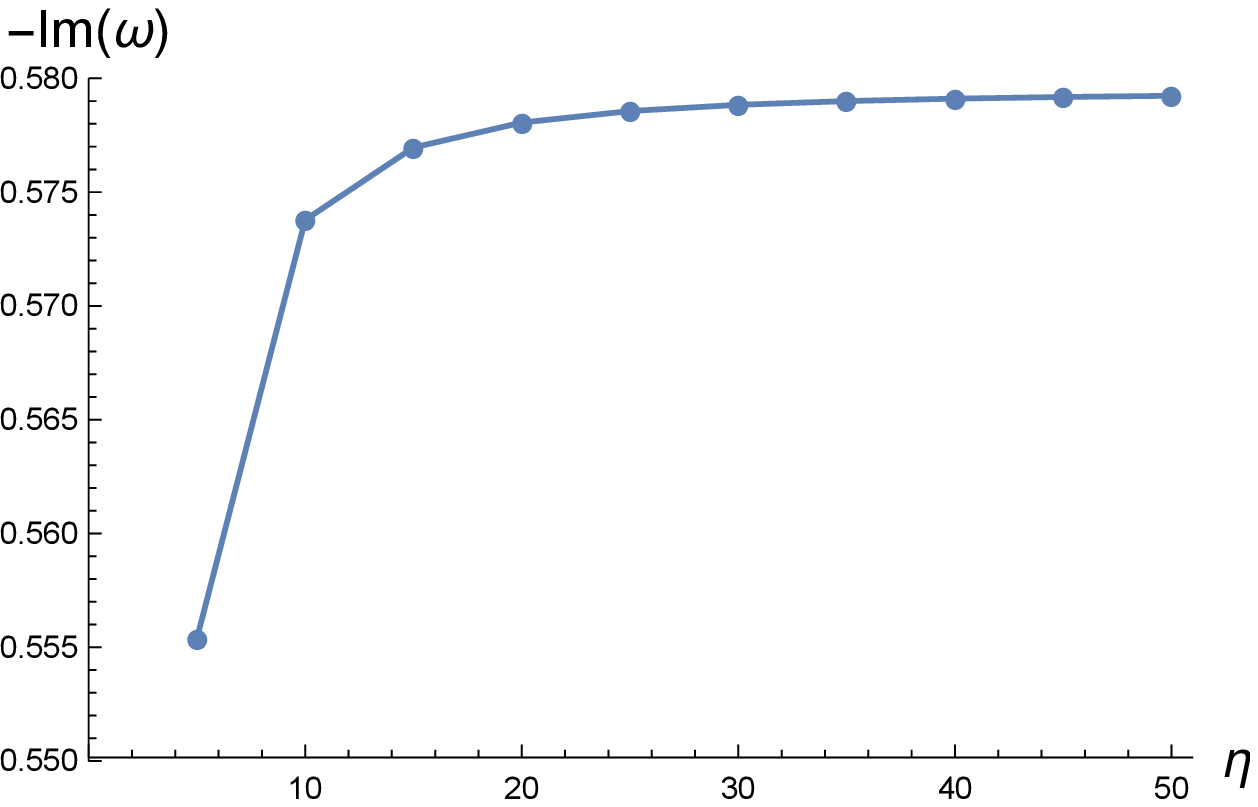}

	\caption{The graph on the left shows the real part of frequencies with respect to the refractive index and on the right the negative imaginary part with respect to the refractive index for the case of $l=3$ and $n=1$.}
	\label{Re&xu}
\end{figure}

We turn to the evolution behavior of real and imaginary parts with respect to an increasing refractive index. We find that the behavior is similar to the cases of different $l$ and $n$ under the condition $l\ge n$.
For example, we investigate the evolution for the case of $l=3$ and $n=1$ by using the Borel-Pad$\acute{\text e}$ summation and observe the behaviors of real and negative imaginary parts as shown in Figure{~\ref{Re&xu}}.

\begin{figure}[H]
    \centering
    
    \includegraphics[width=90mm]{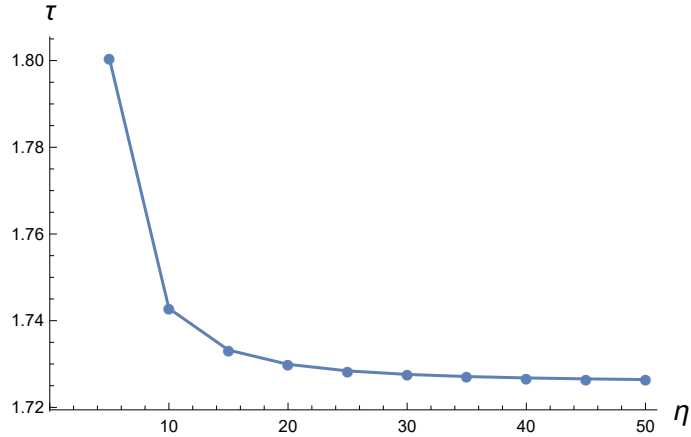}

	\caption{The damping time for the case of $l=3$ and $n=1$.}
	\label{tau}
\end{figure}

It is obvious to see from Figure~\ref{Re&xu} that the real part of frequencies is decreasing when the refractive index increases, whereas the negative imaginary part is changing in an opposite tendency. We plot the dependence of the damping time on the refractive index $\eta$ in Figure~\ref{tau}. Based on the results depicted by Figure~\ref{Re&xu} and Figure~\ref{tau} for the cases of low overtone numbers, i.e. $n\le l$, we point out that an optical black hole in a high refractive index medium oscillates at a low frequency and a short damping time $\tau=\rvert\text{Im}(\omega)\rvert^{-1}$.

\subsection{Behavior at a high overtone number}
\begin{figure}[H]
	\centering
		\includegraphics[width=90mm]{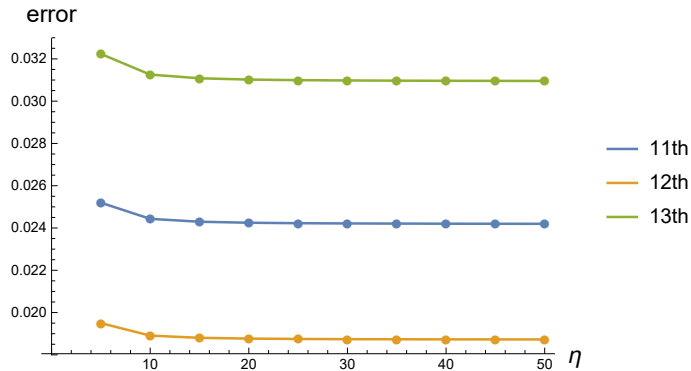}
	
	\caption{The dependence of the error estimates of the 11th, 12th and 13th order WKB-Pad\'e approximations on the refractive index $\eta$ for the case of $l=2$ and $n=8$.}
	\label{err}
\end{figure}

In general, one can compute the quasinormal mode frequencies accurately enough at low overtone numbers, i.e., $n\leq l$, by using approximation methods. However, the computation will be inaccurate and unstable at high overtone numbers, $n\ge l$.
Here we mention that the WKB-Pad\'e approach can suitably be applied to our model at a high overtone number, that is, the errors can effectively be reduced in this case.
To this end, we calculate the error estimates of frequencies from the first to thirteenth order approximation by following ref.~\cite{KZZ} and find that they indeed decrease when the refractive index increases, see Figure~\ref{err} for the 11th, 12th and 13th order WKB-Pad\'e approximations as samples. This feature provides us a possibility to analyze the behaviors of optical black holes at a high overtone number because it implies that the WKB-Pad\'e method, together with  the Borel-Pad$\acute{\text e}$ summation, is applicable as long as the refractive index is big enough.

\begin{figure}[H]
	\begin{center}
		\includegraphics[width=90mm]{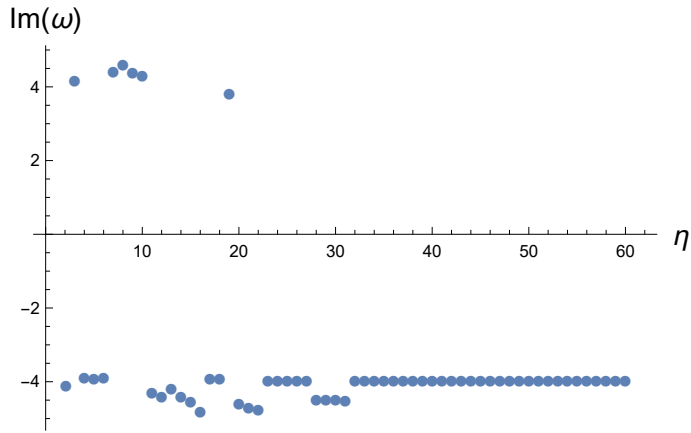}
	\end{center}
	\caption{The discrete points represent the imaginary parts of frequencies by the Borel-Pad$\acute{\text e}$ summation for the case of $l=2$ and $n=8$.}
	\label{Im}
\end{figure}

The imaginary parts of frequencies are usually negative at low overtone numbers, which means that the system of (optical) black holes is stable. But the imaginary parts of frequencies change their signs at high overtone numbers. In accordance with the feature mentioned above, 
we investigate the change of signs of the imaginary parts of frequencies with respect to the refractive indices by using  the Borel-Pad$\acute{\text e}$ summation, and plot the relations between the imaginary parts and different refractive indices in Figure~\ref{Im}.

It is clear that the imaginary parts change their signs from positive to negative, and vice versa when the indices of refraction are not big enough, that is, $\eta < 31$. Nonetheless, they maintain a stable negative value when $\eta\ge31$. This implies that the optical black hole undergoes one phase transition from an unstable mode to a stable one and keeps staying at the stable mode when the refractive index becomes greater than and equal to 31.

\section{Summary}\label{sec4}

We compute numerically the quasinormal mode frequencies of optical black holes in moving dielectrics and give the range of refractive indices required to form a black hole horizon. We find how the refractive index affects the behavior of optical black holes. In the case of low overtone numbers, the optical black hole formed in a high refractive index medium oscillates at a low frequency and returns to a stable status in a short time after being perturbed. In the case of high overtone numbers, the optical black hole will undergo one phase transition from an unstable mode to a stable one and keep staying at the stable mode 
when the refractive index is greater than and equal to 31. In particular, the medium with negative refractive indices plays the same role as the medium with positive refractive indices in our model of optical black holes.
At last, we expect the tests of the phase transition at high overtone numbers and of the mono-effect in both the positive and negative refractive index media in lab of optics in the near future.

\section*{Acknowledgments}
The authors would like to thank Yun-Gui Gong, Yong-Ge Ma, Qi-Yuan Pan, Hao Yang, and Liu Zhao for the enlightening discussions, and in particular the anonymous referee for the helpful comments that improve this work greatly. This work was supported in part by the National Natural Science Foundation of China under grant No.11675081.

\end{document}